\documentclass{article}

\usepackage[final]{neurips_2024}

\usepackage[utf8]{inputenc} 
\usepackage[T1]{fontenc}    
\usepackage{hyperref}       
\usepackage{url}            
\usepackage{booktabs}       
\usepackage{amsmath}        
\usepackage{amsfonts}       
\usepackage{nicefrac}       
\usepackage{microtype}      
\usepackage{xcolor}         
\usepackage{graphicx}

\title{NLCG-Net: A Model-Based Zero-Shot Learning Framework for Undersampled Quantitative MRI Reconstruction}

\author{%
\textbf{Xinrui Jiang}$^1$, \textbf{Yohan Jun}$^{2,3}$, \textbf{Jaejin Cho}$^{2,3}$, \textbf{Mengze Gao}$^4$,\\
\textbf{Xingwang Yong}$^5$, \textbf{Berkin Bilgic}$^{2,3}$\thanks{Correspondence to: \texttt{bbilgic@mgh.harvard.edu}.}\\
$^1$Fudan University\quad
$^2$Harvard Medical School\quad
$^3$Massachusetts General Hospital\\
$^4$Stanford University\quad
$^5$Zhejiang University\\
}

\begin{document}

\maketitle

\begin{abstract}
    Typical quantitative MRI (qMRI) methods estimate parameter maps in a two-step pipeline that first reconstructs images from undersampled k-space data and then performs model fitting, which is prone to biases and error propagation. We propose NLCG-Net, a model-based nonlinear conjugate gradient (NLCG) framework for joint T2/T1 estimation that incorporates a U-Net regularizer trained in a scan-specific, zero-shot fashion. The method directly estimates qMRI maps from undersampled k-space using mono-exponential signal modeling with scan-specific neural network regularization, enabling high-fidelity T1 and T2 mapping. Experimental results on T2 and T1 mapping demonstrate that NLCG-Net improves estimation quality over subspace reconstruction at high acceleration factors.

\end{abstract}

\section{Introduction}
\label{intro}

Standard quantitative MRI (qMRI) techniques rely on a two step process whereby undersampled k-space data are reconstructed first, and then used in model fitting to estimate parameters of interest. Model based approaches \cite{Sumpf2011ModelBasedT2,Wang2018ModelBasedT1} have been developed to incorporate mono-exponential signal models into the reconstruction, so that parameter maps can be directly estimated from undersampled data.

In this work, we propose a Nonlinear Conjugate Gradient (NLCG) optimization to solve the arising optimization problem and use a scan-specific Neural Network as regularizer. Experiments show the ability of the proposed NLCG-Net to improve T1 and T2 mapping relative to subspace modeling at high accelerations, while obviating the need for an external training dataset.

\section{Methods}
\label{methods}

\subsection{Problem Formulation}
We formulate qMRI reconstruction as an optimization problem via the following objective function:

\begin{equation}
  \arg\min_{\vec{x}} \left\| \mathbf{PFCM}(\vec{x}) - \vec{y} \right\|^2
  + \lambda \left\| \vec{x} - \vec{z} \right\|^2
  \label{eq:objective}
\end{equation}

where 

\begin{equation*}
  \vec{x} =
  \begin{bmatrix}
    M_x \\
    M_y \\
    R
  \end{bmatrix}
  =
  \begin{bmatrix}
    M_x \\
    M_y \\
    \dfrac{1}{T}
  \end{bmatrix}
\end{equation*}

\begin{figure}[htbp]
  \centering
  \includegraphics[width=\linewidth]{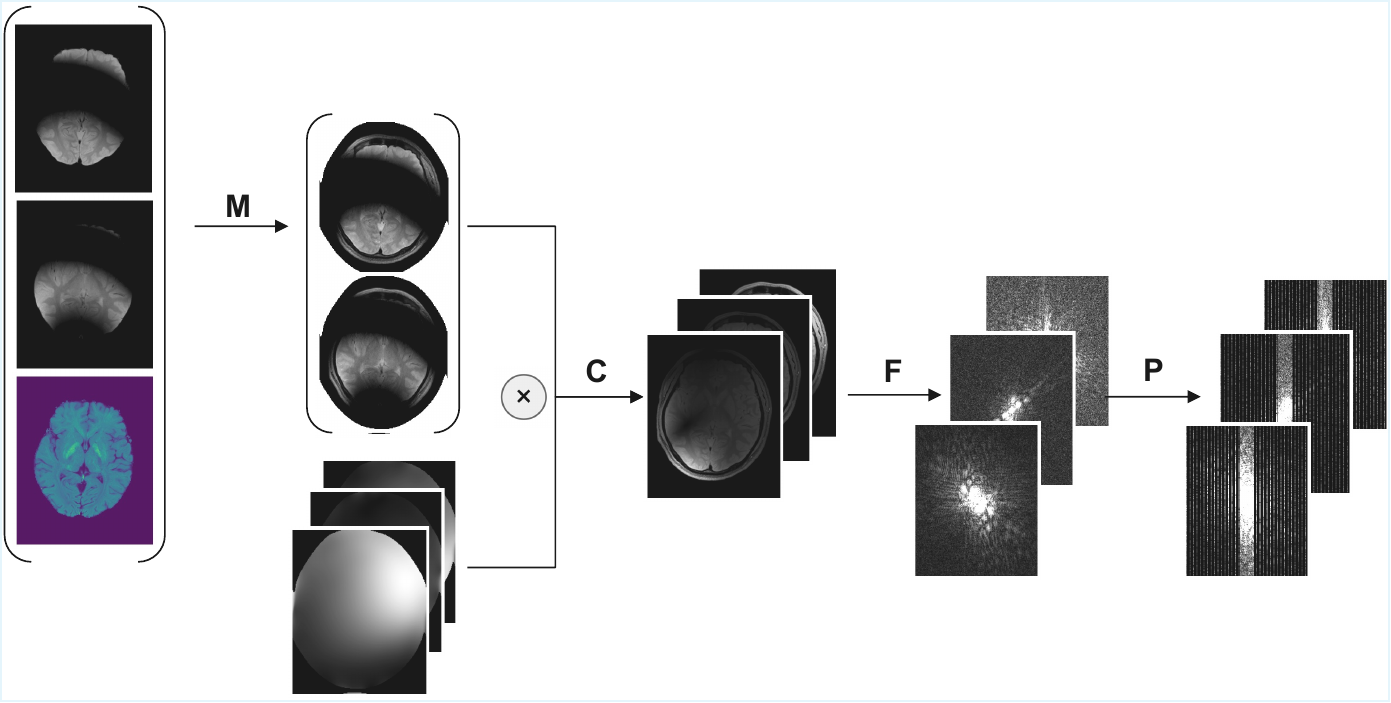}
  \caption{$\mathbf{PFCM}$ forward operators. To derive the k-space expression, $\vec{x}$ is firstly converted to signal intensity following T2/T1 quality, then multiplied with coil sensitivity maps to form each coil image. After that, Fast Fourier Transform $\mathbf{F}$ is performed to transform data into k-space, then finally applying mask $\mathbf{P}$ to reproduce downsample process.}
  \label{fig:pfc_m}
\end{figure}

Here $M_x$, $M_y$ real and imaging components of transverse magnetization, $R$ refers to $R_1$ or $R_2$ representing parameter values, $\vec{y}$ denotes acquired k-space data, $\vec{z}$ refers to regularized $\vec{x}$ and $\lambda$ is regularization coefficient. $\mathbf{PFCM}$ are forward operators illustrated in Fig.~\ref{fig:pfc_m}. $\mathbf{P}$ denotes k-space sampling mask, $\mathbf{F}$ denotes Fast Fourier transform, $\mathbf{C}$ denotes coil sensitivity maps, and $\mathbf{M}$ denotes the mono-exponential signal model, which has different expressions for T2 and T1 mapping:

\begin{subequations}
\label{eq:M_T}
\begin{equation}
  \mathbf{M}_{\mathrm{T2}}:
  \vec{x} =
  \begin{bmatrix}
    M_x\\
    M_y\\
    R_2
  \end{bmatrix}
  \longmapsto
  \begin{bmatrix}
    M_x e^{-\mathrm{TE}\, R_2}\\
    M_y e^{-\mathrm{TE}\, R_2}
  \end{bmatrix},
  \label{eq:M_T2}
\end{equation}
\begin{equation}
  \mathbf{M}_{\mathrm{T1}}:
  \vec{x} =
  \begin{bmatrix}
    M_x\\
    M_y\\
    R_1
  \end{bmatrix}
  \longmapsto
  \begin{bmatrix}
    M_x \bigl(1 - 2 e^{-\mathrm{TI}\, R_1}\bigr)\\
    M_y \bigl(1 - 2 e^{-\mathrm{TI}\, R_1}\bigr)
  \end{bmatrix},
  \label{eq:M_T1}
\end{equation}
\end{subequations}

To solve this, we can unroll the target function and solve it separately and iteratively. Specifically~\cite{Aggarwal2019MoDL},

\begin{subequations}
\begin{equation}
  \vec{z}_n = \mathbf{R}(\vec{x}_n)
  \label{eq:z_update}
\end{equation}
\vspace{-0.3em}
\begin{equation}
  \vec{x}_{n+1} = \arg\min_{\vec{x}} \left\| \mathbf{PFCM}(\vec{x}) - \vec{y} \right\|^2
  + \lambda \left\| \vec{x} - \vec{z}_n \right\|^2
  \label{eq:x_update}
\end{equation}
\end{subequations}

Where $\mathbf{R}$ refers to regularization operation by Neural Network.

\subsection{Proposed Model}
\begin{figure}[t]
  \centering
  \includegraphics[width=\linewidth]{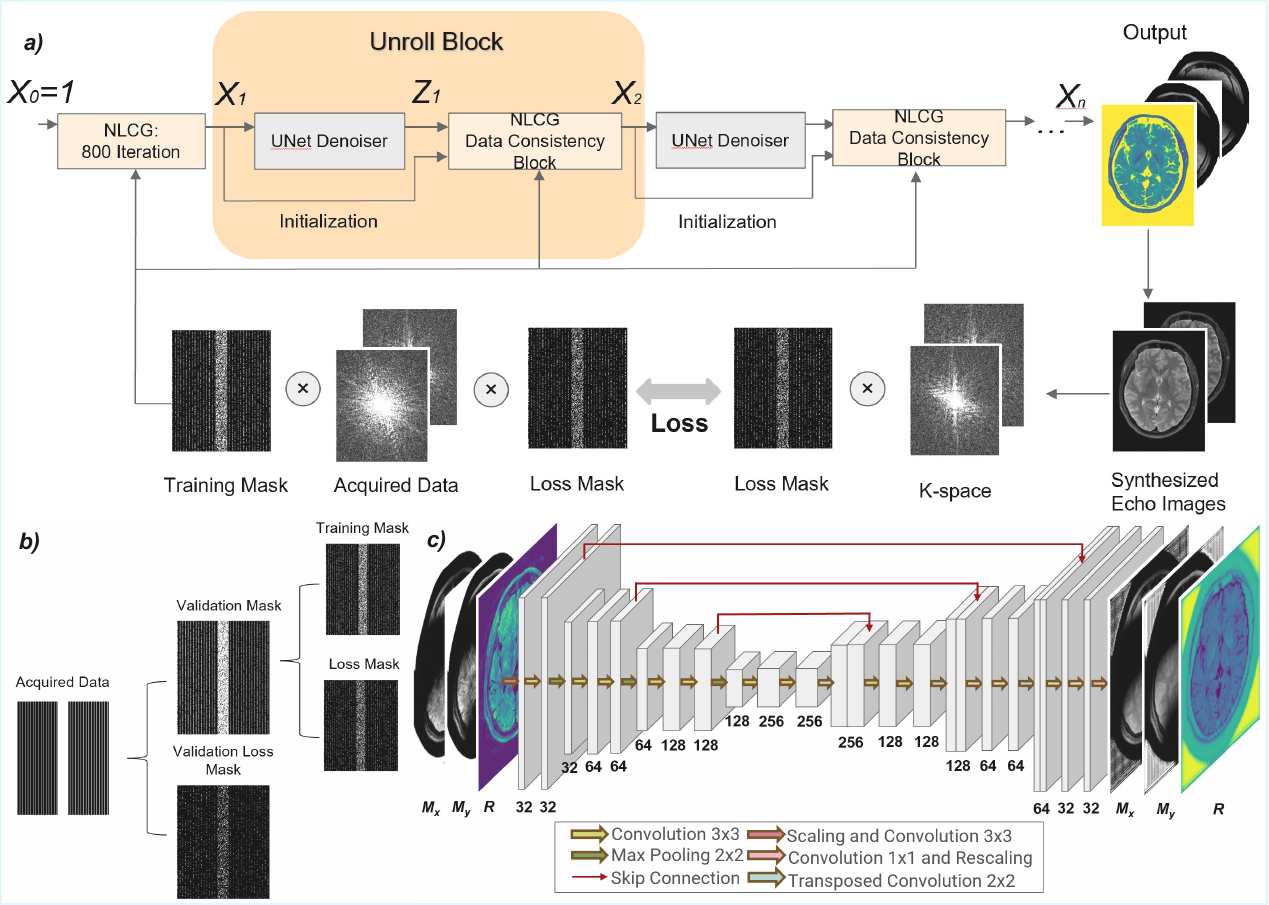}
  \caption{a) NLCG-Net framework. The model performs 800 NLCG iterations first to initialize, then passes to Unroll blocks. A NLCG data consistency layer and U-Net are deployed
  in each block to perform iterative optimization. b) Self-supervised training strategy for
  NLCG-Net. Acquired data are divided into training and validation sets by masking.
  c) NLCG-Net deploys a light U-Net model with fewer convolutional layers and jointly regularizes
  all desired maps.}
  \label{fig:framework}
\end{figure}

We propose NLCG-Net to solve the iterative optimization problem. The detailed architecture of NLCG-Net is presented in Fig.~\ref{fig:framework}. NLCG-Net directly takes k-space data
as input, and unrolls the optimization into several unroll blocks. Each block has one regularization layer and one data consistency (DC), which correspond to Eqs.~\eqref{eq:z_update} and~\eqref{eq:x_update}. For the regularizer, we deploy a light U-Net model with only three downsample and upsample layers to restrict the parameter number and facilitate training. In the data consistency layer, we deploy NLCG to solve Eq.~\eqref{eq:x_update}.

\subsection{Implementation Details}
NLCG-Net contains 3 Unroll blocks. A DC layer performs 20 iterations with NLCG and takes $\vec{x}$ from the last layer as iteration initialization. In the beginning, input $\vec{x}$ is acquired by performing NLCG without regularization to k-space input for 800 iterations, which can be done within 1
minute. To achieve zero-shot training, we use different sampling masks to divide data into training and validation sets, and the splitting strategy is the same as \cite{Yaman2021ZeroShot}. Before feeding into the network, $M_x$, $M_y$, and $R$ are scaled~\cite{Blumenthal2024NLINVNet,Jun2023SSLQALAS} 
by three trainable parameters due to their different dynamic ranges. 
Overall training process was performed on an NVIDIA 4090 GPU, and 300 epochs were finished within 5 hours.

\subsection{Experiments}
We used a set of fully sampled spin echo multi-coil data with FOV 256 × 208 for T2 mapping. 8 echo data are deployed from TE = 23 ms to 184 ms, with an echo spacing of 23ms. We simulate Acceleration by creating sampling masks, with acceleration factor $R = 4$ and $R = 6$. Central ACS region was used with width = 24.

We also validated T1 mapping on in vivo data, which has been prospectively accelerated by $R = 2$. 5 sets of inversion recovery TSE data with TI from 35 ms to 3000 ms were acquired. We
furthermore simulated the $R = 4$ condition without utilizing ACS region data for reconstruction. For metric consideration, we used normalized root mean squared error (NRMSE).

\section{Results}
\begin{figure}[t]
  \centering
  \includegraphics[width=\linewidth]{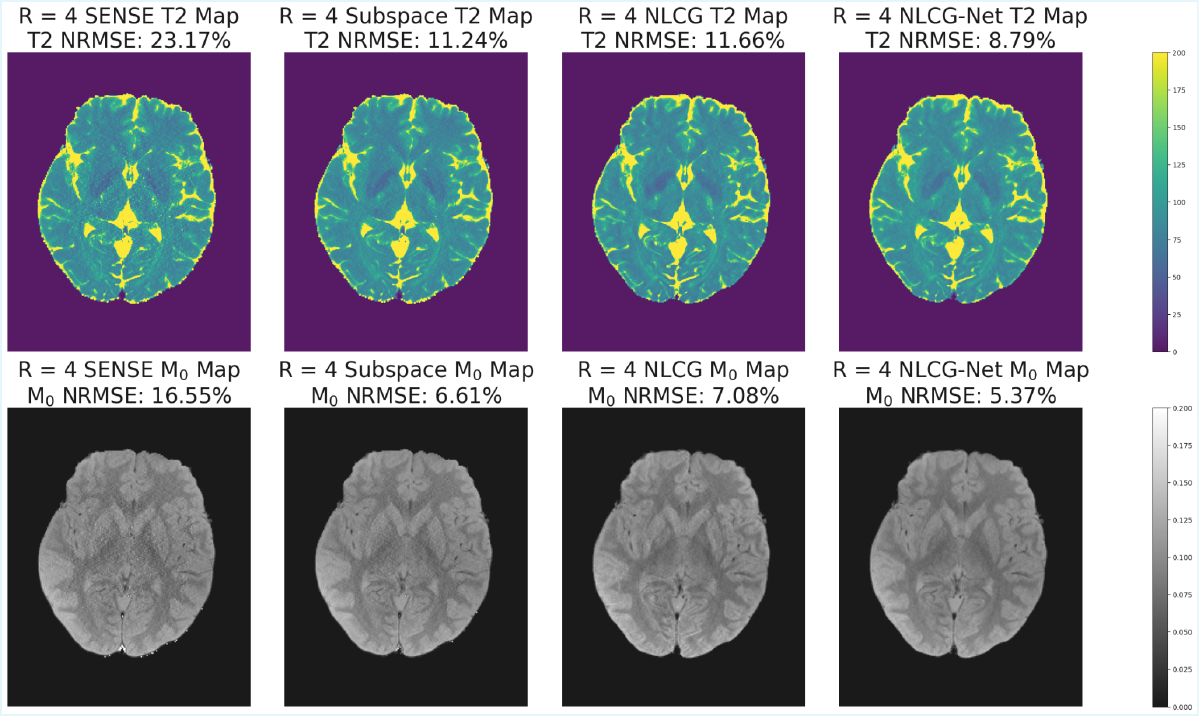}
  \caption{T2 mapping reconstruction results under acceleration rate $R = 4$.}
  \label{fig:t2_r4}
\end{figure}

\begin{figure}[t]
  \centering
  \includegraphics[width=\linewidth]{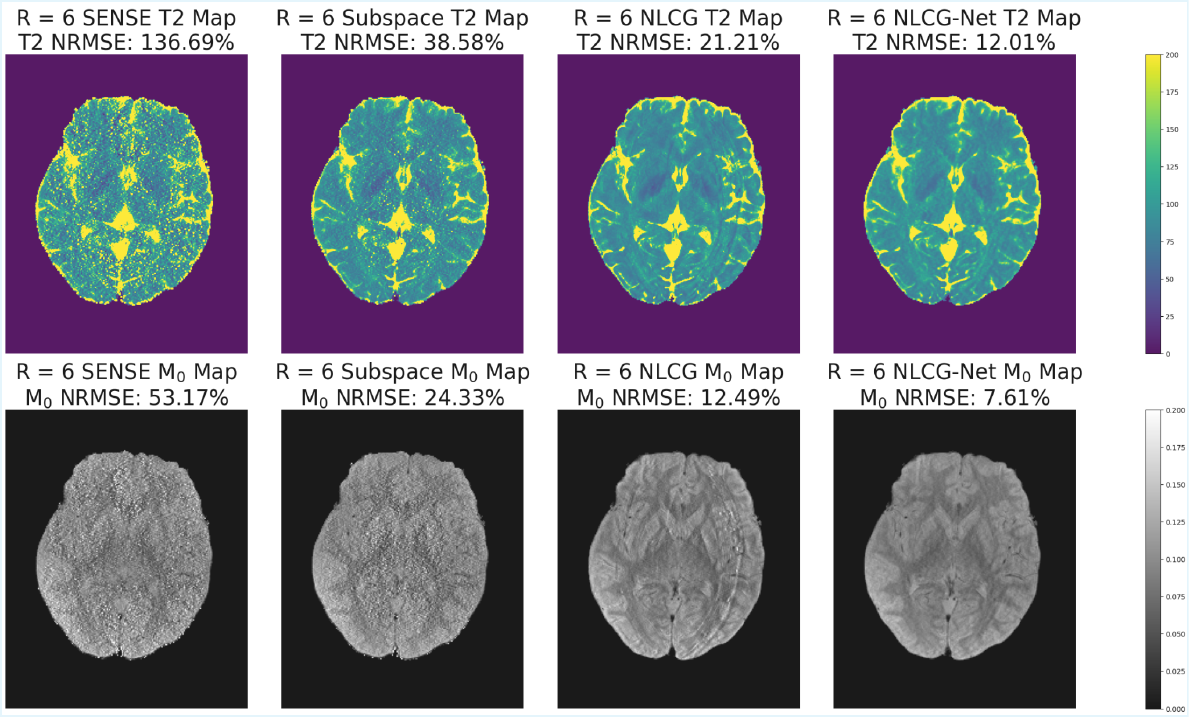}
  \caption{T2 mapping reconstruction results under acceleration rate $R = 6$.}
  \label{fig:t2_r6}
\end{figure}

\begin{figure}[t]
  \centering
  \includegraphics[width=0.95\linewidth]{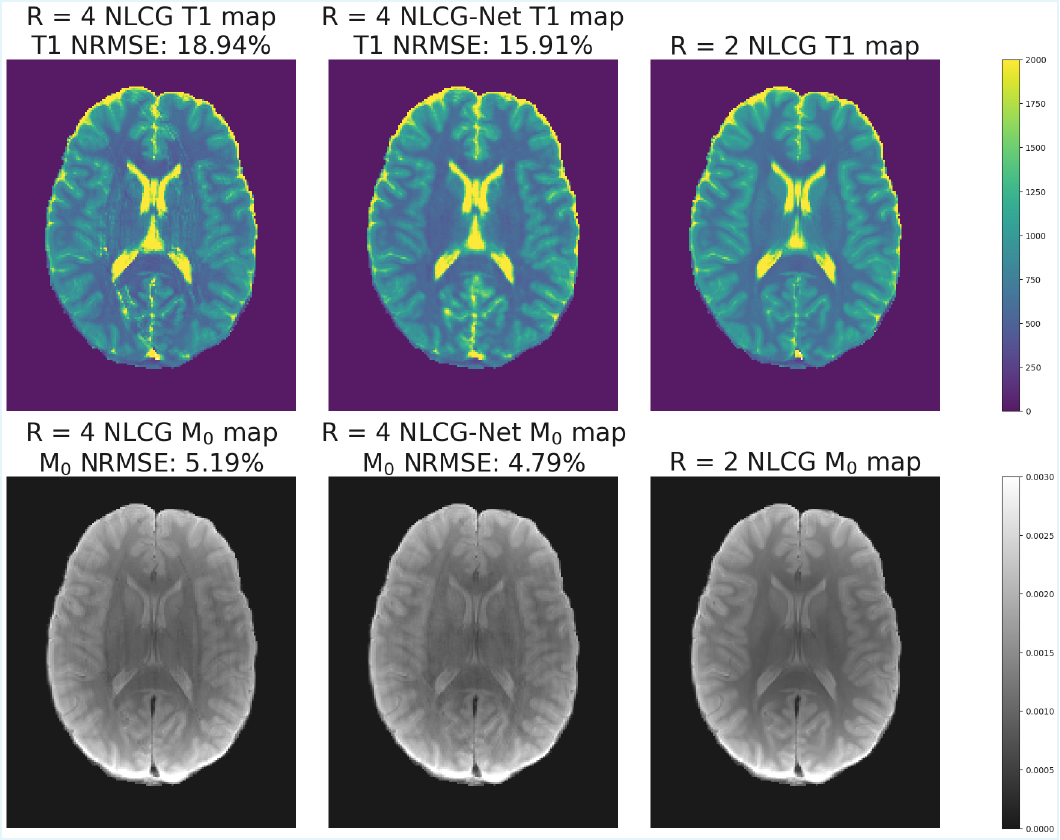}
  \caption{T1 mapping reconstruction results under acceleration rate $R = 4$.}
  \label{fig:t1_r4}
\end{figure}

We present T2, T1, and corresponding M0 map reconstruction results in Figs.~\ref{fig:t2_r4}--\ref{fig:t1_r4}. 
For comparison, we also present T2 estimation using SENSE and Subspace. We observe that for T2 mapping, unregularized NLCG can readily achieve a good fit when considering NRMSE. However, as $R$ increases, aliasing artifacts emerge, which are better mitigated using NLCG-Net. The proposed model retains the lowest NRMSE and effectively suppresses artifacts. For T1 mapping condition, it is observed that NLCG-Net has a similar performance.

\section{Discussion and Conclusion}
We proposed a model-based zero-shot self-supervised learning framework, NLCG-Net for qMRI reconstruction, which is able to reach high acceleration factors with high fidelity. Its nonlinear
estimation is flexible enough for both T2 and T1 mapping, and iterative optimization formulation allows neural network regularization while obviating the need for an external training
datasets.

\bibliographystyle{plain}
\bibliography{refs}

\end{document}